\documentclass[twocolumn,aps,superscriptaddress,showpacs]{revtex4}

\usepackage{epsfig,bm,feynmf}

\usepackage{graphics}

\begin{document}



\title{$J/\psi$ production and elliptic flow in relativistic heavy-ion collisions}


\author{Taesoo Song}\email{songtsoo@yonsei.ac.kr}
\affiliation{Cyclotron Institute, Texas A$\&$M University, College Station, TX 77843-3366, USA}
\author{Che Ming Ko}\email{ko@comp.tamu.edu}
\affiliation{Cyclotron Institute and Department of Physics and Astronomy, Texas A$\&$M University, College Station, TX 77843-3366, USA}
\author{Su Houng Lee}\email{suhoung@yonsei.ac.kr}
\affiliation{Institute of Physics and Applied Physics, Yonsei University, Seoul 120-749, Korea}
\author{Jun Xu}\email{xujun@comp.tamu.edu}
\affiliation{Cyclotron Institute, Texas A$\&$M University, College Station, TX 77843-3366, USA}


\begin{abstract}
Using a two-component model for charmonium production, which includes contributions from both the initial hard nucleon-nucleon scattering and from the regeneration in the quark-gluon plasma, we study the nuclear modification factor $R_{AA}$ and elliptic flow $v_2$ of $J/\psi$ in relativistic heavy ion collisions. For the expansion dynamics of produced hot dense matter, we introduce a schematic fireball model with its transverse acceleration determined from the pressure gradient inside the fireball and azimuthally anisotropic expansion parameterized to reproduce measured $v_2$ of light hadrons. We assume that light hadrons freeze out at the temperature of 120 MeV while charmonia at 160 MeV, similar to the kinetic and chemical freeze-out temperatures in the statistical model, respectively. For the properties of charmonia in the quark-gluon plasma, we use the screening mass between their charm and anticharm quarks and their dissociation cross sections given by the perturbative QCD (pQCD) in the leading order and up to the next-to-leading order, respectively. For the relaxation time of charm and anticharm quarks in the quark-gluon plasma, we also use the one calculated  in the leading order of pQCD. Modeling the effect of higher-order corrections in pQCD by introducing multiplicative factors to the dissociation cross section of charmonia and the elastic scattering cross sections of charm and anticharm quarks, we find that this effect is small for the $R_{AA}$ of $J/\psi$ as they suppress the number of initially produced $J/\psi$ but enhance the number of regenerated ones. The higher-order corrections increase, however, the $v_2$ of $J/\psi$. Our results suggest that the $v_2$ of $J/\psi$ can play an important role in discriminating between $J/\psi$ production from the initial hard collisions and from the regeneration in the quark-gluon plasma.
\end{abstract}

\pacs{} \keywords{}

\maketitle


\section{Introduction}

One of the signatures suggested for the formation of the quark-gluon plasma (QGP) in relativistic heavy ion collisions is the suppressed production of $J/\psi$~\cite{Matsui:1986dk}. This was based on the idea that once the QGP is formed in heavy-ion collisions, the $J/\psi$ produced from initial hard scattering would be dissociated in the QGP because of the color screening between its charm and anticharm quark, reducing thus the number of produced $J/\psi$. However, recent results from Lattice Quantum Chromodynamics (LQCD) calculations have suggested that $J/\psi$ can exist above the critical temperature for the phase transition between the QGP and the hadron gas (HG)~\cite{Hatsuda04,Datta04}. This thus not only reduces the dissociation of $J/\psi$ but also can lead to the production of $J/\psi$ from the recombination of charm  and anticharm quarks in QGP~\cite{Thews:2000rj}. The latter effect becomes more important when the collision energy is high as more charm and anticharm quarks are produced from initial hard scattering.

Experimentally, suppressed production of $J/\psi$ in heavy-ion collisions compared with that is expected from initial hard scattering has been observed at both Super Proton Synchrotron (SPS) and Relativistic Heavy Ion Collider (RHIC)
\cite{Abreu:1997jh,Arnaldi:2007aa,Adare:2006ns,Adler:2005ph}. Many theoretical explanations have been proposed to understand these experimental observations~\cite{Zhang:2000nc,Grandchamp:2002wp,Yan:2006ve,Andronic:2006ky}. Among them are the two-component model~\cite{Grandchamp:2002wp} and the statistical model~\cite{Andronic:2006ky}.
In the two-component model, some of the initially produced $J/\psi$ were found to survive during the collision if the $J/\psi$ dissociation temperature is sufficiently high~\cite{Grandchamp:2002wp}. In the statistical model, it is assumed, however, that all initially produced $J/\psi$ are dissociated in QGP and the detected $J/\psi$ are entirely due to regeneration from charm and anticharm quarks in the QGP~\cite{Andronic:2006ky}.
Both models have been able to successfully reproduce
the measured nuclear modification factor $R_{AA}$ of $J/\psi$ as a function of the number
of participants~\cite{Andronic:2006ky,Grandchamp:2002wp,Song:2010ix}
as well as of the transverse momentum of $J/\psi$~\cite{Akkelin:2009nx,Zhao:2007hh}.
In the two-component model, the number of regenerated $J/\psi$
is smaller than that in the statistical model, because not all charm quarks produced
in a collision are thermalized in QGP, the finally produced $J/\psi$ include, however,
those produced from initial hard scattering and having survived thermal decay
in the produced hot dense matter. Although the transverse momentum of
thermally produced $J/\psi$ in the statistical model is smaller than
that of $J/\psi$ produced from initial hard scattering, it is
enhanced by the transverse flow of $J/\psi$. Therefore, it is difficult to discriminate the two models
based on the number of finally produced $J/\psi$ and their transverse momentum distribution.

Previously, two of us have studied in the two-component model the
$R_{AA}$ of $J/\psi$ as a function of the number of participants in
heavy-ion collisions \cite{Song:2010ix}. In that
study, the initial local temperature of produced QGP was determined
from the local entropy density, which was taken to be proportional to the linear
combination of the number of participants and the number of binary collisions in
that region. The initial fireball thus separates into two regions with temperatures higher and lower than the dissociation temperature of $J/\psi$. While the survival rate of initially
produced $J/\psi$ vanishes in the high temperature region, its value in the low temperature region is finite and depends on the temperature. The different values of the $J/\psi$ survival rate in the two regions were found to give rise to a kink in the
dependence of the $R_{AA}$ of $J/\psi$ on the number of participants as seen in the experimental
data. In the above mentioned work, both the screening mass between
charm and anticharm quark and the relaxation factor of charm quarks in QGP
were calculated in the leading order in pQCD, while the dissociation cross section of
charmonia in QGP were calculated up to the next-to-leading order in pQCD. According to LQCD, the strong coupling constant remains large up to around 2 $T_c$, so below this temperature the QGP is nonperturbative and higher-order corrections in pQCD may not be negligible. The higher-order corrections are expected to enhance in QGP both the thermal decay rate of charmonia, which would decrease the number of initially produced $J/\psi$,
and the relaxation of charm quarks, which would increase the number of regenerated $J/\psi$.
The cancelation of these two effects is probably one of the reasons that both the
two-component model and the statistical model can
reproduce the measured $R_{AA}$ of $J/\psi$ as a function of the number of
participants in heavy-ion collisions.

In the present work, we improve our previous two-component model in
two ways. Firstly, instead of assuming that the transverse
acceleration of produced fireball is constant and then vanishes at a certain time as in the previous work, the acceleration in the present work is obtained from the pressure of the fireball and
continues to be present until the fireball freezes out.
Secondly, we take into account the azimuthally anisotropic expansion
in noncentral collisions by using a simple parametrization
with a transverse-momentum-dependent flow velocity to
reproduce the measured $v_2$ of light hadrons. This schematic
fireball expansion model is then applied to study the $R_{AA}$ of
$J/\psi$ as a function of both the number of participants and the transverse momentum of $J/\psi$ as well as to study the $v_2$ of $J/\psi$. For the chemical and thermal freeze-out temperatures, we follow those used in the statistical model. In this model, the chemical
freeze-out occurs at temperature around 160 MeV, after which
chemical equilibrium is reached and no additional effective chemical
reactions are present and the kinetic freeze-out
takes place later when the temperature drops to 120 MeV. Since the kinetic
freeze-out depends on the elastic scattering cross sections of
hadrons and unlike light hadrons, the elastic scattering cross sections of
$J/\psi$ in hadronic matter are small, we assume that the kinetic freeze-out temperature for $J/\psi$ is the same as the chemical freeze-out temperature. Finally, we take into account the higher-order corrections by multiplying the pQCD cross sections with arbitrary constants to study their effects on the $R_{AA}$ and $v_2$ of $J/\psi$. As shown later, although the fraction of $J/\psi$ produced from regeneration increases with increasing larger higher-order corrections, the resulting $R_{AA}$ of $J/\psi$ remains essentially unchanged. On the other hand, the $v_2$ of  $J/\psi$ at low transverse momentum becomes more negative with larger higher-order corrections. Therefore, studying the $v_2$ of $J/\psi$ can help to discriminate between the two $J/\psi$ production mechanisms of from initial hard scattering and regeneration in the QGP.

This paper is organized as follows. In Sec.~\ref{two}, we briefly describe the two-component model used in Ref.~\cite{Song:2010ix}. We then introduce in Sec.~\ref{fireball} the schematic model for the expanding fireball produced in heavy-ion collision to obtain the time evolution of the thermal properties as well as the spatial and kinematic information of the fireball for both central and non-central collisions. In Sec.~\ref{raa}, the
$R_{AA}$ of $J/\psi$ as a function of the number of participants and
the transverse momentum of $J/\psi$ are calculated with the schematic
model. Both the Cronin effect for the initial transverse
momentum broadening and the leakage effect on initially produced $J/\psi$ are
considered. Results on the elliptic flow of $J/\psi$ are given in
Sec.~\ref{elliptic}. In Sec.~\ref{high}, higher-order corrections
are introduced by multiplying the pQCD results with constant
factors, and their effects on the $R_{AA}$ and $v_2$ of $J/\psi$ are investigated.
Finally, the summary and conclusion are given in Sec.~\ref{summary}.
Also, details on the the azimuthal angle dependence of the flow velocity are
given in the Appendix.

\section{the two-component model for $J/\psi$ production}\label{two}

In the two-component model for $J/\psi$ production in relativistic heavy ion collisions,
some of initially produced $J/\psi$ from hard scattering can survive through the collisions and
contribute to the final observed $J/\psi$ in experiments, besides those regenerated from the charm and anticharm quarks in the produced QGP. In this Section, we briefly describe the method used in Ref~\cite{Song:2010ix} to evaluate these two contributions.

\subsection{directly produced $J/\psi$}

The number of initially produced $J/\psi$ in a heavy ion collision at impact parameter $b$ can be determined by using the Glauber model, that is
\begin{eqnarray}
N^{AA}_{J/\psi}=\sigma_{J/\psi}^{NN}A^2 T_{AA}({\vec b}),
\end{eqnarray}
where $\sigma_{J/\psi}^{NN}$ is the cross section for $J/\psi$ production in a $p+p$ collision, which has a value of 0.774 $\mu{\rm b}$ per rapidity near midrapidity at collision energy $\sqrt{s}=200$ GeV \cite{Adare:2006kf}, and A is the mass number of colliding nuclei. The overlap function $T_{AA}$ in the above equation is defined by
\begin{eqnarray}
T_{AA}=\int d^2s T_A({\vec s})T_A({\vec b}-{\vec s})
\end{eqnarray}
in terms of the thickness function
\begin{eqnarray}
T_A({\vec s})=\int_{-\infty}^{\infty}dz\rho_A({\vec s},z),
\end{eqnarray}
where $\rho_A({\vec s},z)$ is the nucleon distribution in the colliding nucleus.

The survival rate of initially produced $J/\psi$ in relativistic heavy ion collisions is then given by
\begin{eqnarray}
S_{\rm th}({\vec b},{\vec s})=\exp\left\{-\int_{\tau_0}^{\tau_{\rm cf}}\Gamma(\tau^\prime)d\tau^\prime,\right\}
\end{eqnarray}
where $\tau_0$ is the time the $J/\psi$ is produced from initial hard scattering and $\tau_{\rm cf}$ is the time the $J/\psi$ freezes out from the hot dense matter. The $J/\psi$ survival rate depends on its  thermal decay width $\Gamma(T)$ in the produced hot dense matter,
\begin{eqnarray}
\Gamma(T)&=&\sum_i \int\frac{d^3k}{(2\pi)^3}v_{\rm rel}(k)n_i(k,T) \sigma_i^{\rm diss}(k,T).
\end{eqnarray}
In the above, the summation is over the particle species $i$, which includes the quark and gluon in QGP as well as the baryon and meson in HG, with its density denoted by $n_i(k,T)$; $v_{\rm rel}$ is the relative velocity between the $J/\psi$ and the particle; and $\sigma_i^{\rm diss}$ is the dissociation cross section of the $J/\psi$ by the particle.

Since some directly produced $J/\psi$ are from the decay of directly produced
excited charmonium states $\chi_c$ and $\psi^\prime$, the survival rate of all initially produced $J/\psi$ from thermal decay is the linear combination of the survival rates for $J/\psi$, $\chi_c$, and $\psi^\prime$, that is
\begin{eqnarray}
S_{\rm th}({\vec b},{\vec s})&=&0.67 S^{J/\psi}_{\rm th}({\vec b},{\vec s})+0.25 S^{\chi_c}_{\rm th}({\vec b},{\vec s})\nonumber\\
&& +0.08 S^{\psi^\prime}_{\rm th}({\vec b},{\vec s}),
\end{eqnarray}
where we have assumed that 25\% and 8\% of $J/\psi$ are, respectively, from the feed-down of $\chi_c$ and $\psi^\prime$ as usually adopted in the literature \cite{LindenLevy:2009zz}. We note that in Ref~\cite{Song:2010ix} the dissociate cross sections of charmonia in QGP were calculated in pQCD up to the next-to-leading order and those in HG were calculated via the factorization formula with the Bethe-Salpeter amplitude for the bound charmonium states~\cite{Park:2007zza,Song:2007gm}.

\subsection{regenerated $J/\psi$}

For regenerated $J/\psi$, their number is similarly determined as in the statistical model, that is
\begin{eqnarray}\label{reg}
N^{AA}_{{\rm reg}J/\psi}&=&\gamma^2\left\{n_{J/\psi}S^{J/\psi}_{{\rm th}-H}+Br(\chi_c\to J/\psi)n_{\chi_c}S^{\chi_c}_{{\rm th}-H}\right.\nonumber\\
&&\left.+Br(\psi^\prime\to J/\psi)n_{\psi^\prime}S^{\psi^\prime}_{{\rm th}-H}\right\}VR.
\end{eqnarray}
In the above, $n_{J/\psi}$, $n_{\chi_c}$, and $n_{\psi^\prime}$ are, respectively, the equilibrium densities of $J/\psi$, $\chi_c$, and $\psi^\prime$ at chemical freeze-out when the system has volume $V$. The charm fugacity parameter $\gamma$ takes into account the fact that the total charm quark number is not in chemical equilibrium. It is determined instead by
\begin{eqnarray}
N^{AA}_{c\bar c}=\left\{\frac{1}{2}\gamma n_0\frac{I_1(\gamma n_0V)}{I_0(\gamma n_0V)}+\gamma^2n_h\right\}V,
\end{eqnarray}
where $I_0$ and $I_1$ are modified Bessel functions with $n_0$ and $n_h$ being, respectively, the density of open and hidden charms, and $N^{AA}_{c\bar c}$ is the total charm quark pairs produced from initial hard scattering given by
\begin{eqnarray}
N^{AA}_{c\bar c}=\sigma_{c\bar c}^{NN}A^2 T_{AA}({\vec b}),
\end{eqnarray}
where $\sigma_{c\bar c}^{NN}$ is the cross section for $c\bar c$ pair production in a $p+p$ collision, which is 63.7 $\mu$b per rapidity near midrapidity at $\sqrt{s}=200$ GeV in perturbative QCD \cite{Cacciari:2005rk}.
The factor $R$ in Eq.~(\ref{reg}) takes into account the nonequilibrium effect of charm quarks in QGP and is evaluated according to
\begin{eqnarray}
R=1-\exp\left\{-\int_{\tau_0}^{\tau_{\rm QGP}} d\tau\Gamma_c(T(\tau))\right\},
\end{eqnarray}
where $\tau_0$ and $\tau_{\rm QGP}$ are, respectively, the time charm quark pairs are produced from initial hard scattering and the time QGP phase ends, and $\Gamma_c(T)$ is the thermal scattering width of charm quarks in QGP and is calculated in Ref~\cite{Song:2010ix} using the leading-order pQCD.

\section{A schematic model for fireball expansion}\label{fireball}

In our previous work \cite{Song:2010ix}, it was assumed for
simplicity that the fireball formed in heavy-ion collisions
expands transversely with constant acceleration of 0.1 $c^2$/fm
until certain time. Besides, the expansion is assumed to be azimuthally isotropic even in noncentral
collisions, neglecting thus the initial elliptic shape of the
collision geometry. Since the transverse acceleration
of the fireball is expected to be a smooth and continuous function of time and
also to be azimuthally anisotropic in non-central collisions, we improve the treatment of the acceleration of the fireball by using the thermodynamical consideration and also introduce a simple
parametrization with a transverse-momentum-dependent flow velocity for the anisotropic expansion.

\subsection{Central collisions}

To treat dynamically the transverse acceleration of the fireball formed in relativistic
heavy ion collisions, we consider the pressure inside the fireball. In terms of
the energy density $e$, entropy density $s$, and temperature $T$ of the fireball as well as the chemical potential $\mu_i$ and number density $n_i$ of particle species $i$ in the fireball at a certain time, the pressure $p$ is given by
\begin{equation}\label{pressure}
p=sT-e+\sum_i \mu_i n_i.
\end{equation}
The outward force due to the pressure is then
\begin{equation}
F=pA,
\end{equation}
where $A$ is the area of the cylindrical side of the fireball.
Letting $M$ be the inertia mass against the transverse expansion of
the fireball, the transverse acceleration is then
\begin{equation}
a_T^{}=\frac{(p-p_f)A}{M}, \label{acc}
\end{equation}
where $p_f$ is the pressure of the fireball at freeze-out, if we take
into consideration of the fact that the acceleration should vanish at freeze-out, as particles undergo only free streaming afterwards.
It is assumed in the above that the fireball is globally in thermal
equilibrium. We have thus ignored the fact that in more realistic hydrodynamic approach only local thermal equilibrium is achieved and each fluid element
of the fireball moves collectively.
Assuming the transverse acceleration $a_T$ acts in the local
frame, the transverse velocity of the boundary of the fireball in
the laboratory frame at "i+1"th time step can then be
obtained from its value at "i"th time step by adding the effect of
acceleration relativistically, i.e.,
\begin{equation}
v^{i+1}=\frac{v^i+\Delta v}{1+v^i \Delta v},
\end{equation}
where $\Delta v=a_T^i \Delta t$.

\begin{figure}[h]
\centerline{
\includegraphics[width=8cm]{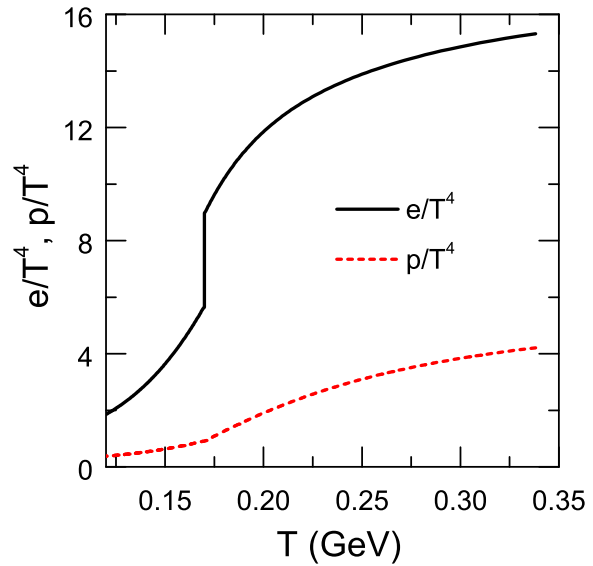}}
\caption{(color online) Energy density (solid line) and
pressure (dashed line) divided by temperature to the forth power in
the quasiparticle model for QGP and the resonance gas model for HG.}
\label{ep}
\end{figure}

For the thermal quantities in Eq.~(\ref{pressure}), we use
the quasiparticle picture for the QGP, which assumes that strongly interacting
massless partons (quarks and gluons) can be substituted by
noninteracting massive partons to reproduce the energy density and pressure
extracted from LQCD. In this approach, the pressure, energy density,
and entropy density are given, respectively, by~\cite{Levai:1997yx}
\begin{eqnarray}
p(T)&=& \sum_i \frac{g_i}{6\pi^2}\int^\infty_0 dk f_i(T) \frac{k^4}{E_i}-B(T)\nonumber\\
&\equiv& p_0(T)-B(T)\nonumber\\
e(T)&=& \sum_i \frac{g_i}{2\pi^2}\int^\infty_0 dk k^2 f_i(T) E_i +B(T)\nonumber\\
s(T)&=& \sum_i \frac{g_i}{2\pi^2T}\int^\infty_0 dk f_i(T) \frac{\frac{4}{3}k^2+m_i^2(T)}{E_i}.
\end{eqnarray}
In the above, $m_i(T)$, $g_i$, and $\mu_i$ are, respectively, the thermal mass, degeneracy factor
and chemical potential of parton species $i$ with $\mu_i$ obtained from the flavor conservation~\cite{BraunMunzinger:1999qy}. The parton distribution function is denoted by
\begin{equation}
f_i(T)=\frac{1}{\exp[(E_i-\mu_i)/T]\pm 1}
\end{equation}
with the plus and minus signs in the denominator for quarks and gluons, respectively,
and $E_i=(m_i^2+k^2)^{1/2}$. For the bag pressure
$B(T)$, it is obtained from the relation $s=\partial p/\partial T$ such that
\begin{eqnarray}
B(T)=B_0 +\sum_i \int_{T_c}^T dT \frac{\partial p_0}{\partial m_i(T)}\frac{\partial m_i(T)}{\partial T},
\end{eqnarray}
where $T_c=170$ MeV is the critical temperature for the QGP to HG phase
transition and $B_0$ is the bag pressure at $T_c$ with its value taken to be 0.1 times the energy density of QGP at $T_c$ to make the pressure continuous at $T_c$. From the above thermal quantities, the parton thermal masses can be determined by fitting the QGP equation of state from LQCD as given in details in Ref.~\cite{Song:2010ix}. For thermal quantities in the HG phase, they are calculated in the resonance gas model by including all mesons of masses less than 1.5 GeV and all baryons of masses less than 2.0 GeV, and for simplicity they are assumed to be noninteracting. Figure~\ref{ep} shows the resulting energy density and pressure, divided by temperature to the fourth power, as functions of  temperature. We note that the pressure at the thermal freeze-out temperature 120 MeV is
$7.75\times 10^{-5}~{\rm GeV^4}$.

The inertia mass $M$ in Eq. (\ref{acc}) is a free parameter in our model and is determined by fitting the average transverse momenta of light hadrons at thermal freeze out to the measured data. Assuming particles freeze out at certain proper time, their number is then
\begin{eqnarray}
N&=&\int d\sigma^\mu J_\mu=\int d\sigma^\mu  u_\mu n\nonumber\\
&=&\int \frac{d^3p}{(2\pi)^3} \int d\sigma \cdot u ~e^{-p\cdot u/T}\nonumber\\
&=&\int \frac{d^3p}{(2\pi)^3E} \int d\sigma \cdot p ~e^{-p\cdot u/T},
\label{number}
\end{eqnarray}
where $\sigma^\mu$ and $J_\mu$ are the hypersurface vector of the
fireball and the current density vector of particles,
respectively; $u_\mu$ and $n$ are, respectively, the 4-velocity of the fluid element
and the number density of particles. In obtaining the last step of the above equation, we have
used the Boltzmann distribution for evaluating the number density and the fact that the 4-velocity of
a fluid element is equal to the average velocity of particles in the fluid element.
Ignoring the transverse boost and keeping
only the longitudinal boost, the hypersurface vector becomes
\begin{eqnarray}
d\sigma^\mu&=&(\partial \tau /\partial t, \vec{0}, \partial \tau /\partial z)\tau d\eta rdrd\phi\nonumber\\
&=&(\cosh \eta, \vec{0}, \sinh \eta)\tau d\eta rdrd\phi,
\label{hypersurface}
\end{eqnarray}
where
\begin{eqnarray}
\eta=\frac{1}{2}\ln\frac{t+z}{t-z}\approx \frac{1}{2}\ln\frac{1+v_z}{1-v_z}\nonumber.
\end{eqnarray}
For the 4-velocity $ u_\mu$ of the fluid element, both boosts
are included, with the transverse boost followed by the longitudinal boost, leading to
\begin{eqnarray}
u_\mu=(\cosh \eta \cosh \rho, \cos \phi \sinh \rho, \sin \phi \sinh
\rho, \sinh \eta \cosh \rho ),\nonumber
\label{velocity}
\end{eqnarray}
where $\rho$ is the transverse rapidity defined as
$\rho=\tanh(v_T)$ with $v_T$ being the transverse velocity. The
transverse rapidity of the fluid element is assumed to be proportional to
the radial distance from the center of the fireball, based on the result
from hydrodynamics \cite{Teaney:2001av}. For the 4-momentum $p_\mu$ of a particle, it is
\begin{eqnarray}
p_\mu=(m_T \cosh y, p_T\cos\varphi, p_T\sin\varphi, m_T \sinh y),\nonumber
\label{momentum}
\end{eqnarray}
where $y$ and $m_T=\sqrt{m^2+p_T^2}$ are, respectively, the rapidity
and the transverse energy. In general, three different azimuthal
angles are required: one for the hypersurface vector $\sigma_\mu$;
one for the 4-velocity $u_\mu$ of the fluid element;
and the other for the momentum vector $p_\mu$. In central collisions, the azimuthal
angle for $\sigma_\mu$ is, however, the same as that for $u_\mu$.
This is also the case in noncentral collisions if only the radial
flow is considered and this is assumed in present study.

Substituting Eqs. (\ref{hypersurface}) into Eq. (\ref{number}) leads
to \cite{Fries:2003kq}
\begin{eqnarray}
&&\frac{dN}{dy dp_T^2}=\frac{1}{2(2\pi)^3}\int_0^{2\pi} d\varphi \int d\sigma\cdot p~ e^{-p\cdot u/T}\nonumber\\
&&=\frac{\tau m_T}{2(2\pi)^3}\int dA_T \int_0^{2\pi} d\varphi  ~e^{p_T \sinh \rho \cos(\phi-\varphi)/T}\nonumber\\
&&~~~\times \int_{-\infty}^\infty d\eta \cosh(\eta-y)~e^{-m_T \cosh\rho \cosh(\eta-y)/T}\nonumber\\
&&=\frac{\tau m_T}{(2\pi)^2}\int dA_T I_0\bigg[\frac{p_T \sinh
\rho}{T}\bigg] K_1\bigg[\frac{m_T \cosh \rho}{T}\bigg],
\end{eqnarray}
where $A_T$ is the transverse area, and $I_n$ and $K_n$ are modified
Bessel functions. From Eq.
(\ref{number}), the averaged transverse momentum of particles is given by
\begin{eqnarray}
\langle p_T\rangle=\frac{\int dp_T^2 p_T (dN/dy dp_T^2)}{\int dp_T^2 (dN/dy dp_T^2)}.
\label{pt}
\end{eqnarray}
We find that with the inertia mass $M=570$ GeV,
the transverse velocity of fireball at freeze-out is 0.71, and the average transverse momenta
of pions, kaons and protons are 462, 668 and 953 MeV/$c$, respectively,
which are comparable to measured values of 451$\pm$33, 670$\pm$78 and
949$\pm$85 MeV/$c$ for $\pi^+$, $K^+$ and proton, respectively, in
Au+Au collisions as $\sqrt{s}=200$ GeV and of 0-5\% centrality \cite{Adler:2003cb}.
We note that the measured average transverse momenta of $\pi^-$, $K^-$ and antiproton
have similar values as those for their antiparticles.

\begin{figure}[h]
\centerline{
\includegraphics[width=8cm]{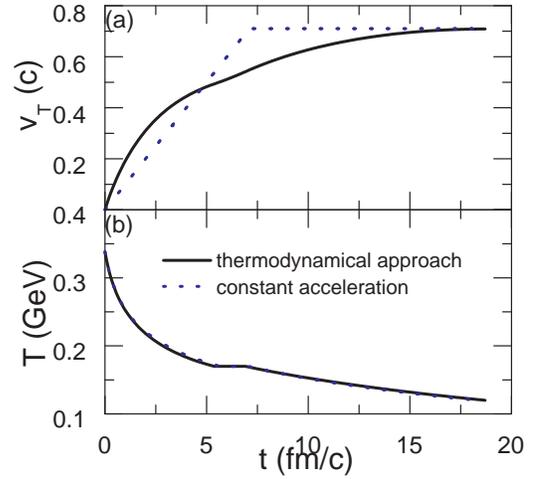}}
\caption{(color online) Transverse velocity (a) and temperature (b) of fireball
in central collision using transverse acceleration from the thermodynamical approach (solid line) or
constant acceleration (dotted line) as functions of time.}
\label{comparison}
\end{figure}

In Fig.~\ref{comparison}(a), we show the transverse velocity of the
fireball obtained with the acceleration using the above
thermodynamical approach as a function of time and compare it with
that using the constant acceleration as in our previous study. The
transverse velocity is seen to increase rapidly at first and then
slowly in the thermodynamical approach, whereas it increases
linearly with time and stays constant after a certain time in our previous study with constant acceleration.
For the temperature of the fireball, shown in
Fig.~\ref{comparison}(b), it evolves in time similarly in the two
cases in spite of their different evolutions in the transverse
velocity. In the thermodynamic approach, the QGP phase terminates at
$t=5.4$~fm/$c$, the mixed phase at $t=6.9$~fm/$c$, and
the HG phase reaches the freeze-out temperature at $t=18.7$~fm/$c$. In
the case of constant acceleration, the corresponding times are
$5.6$~fm/$c$, $7.0$~fm/$c$, and $18.3$~fm/$c$, respectively, and are
thus similar. Therefore, the modified acceleration introduced in the
present study is not expected to change our previous results on the
$R_{AA}$ of $J/\psi$ as a function of the number of
participants~\cite{Song:2010ix}.

\subsection{Non-central collisions}

In noncentral collisions, the length of the overlapping region of two
colliding nuclei is shorter in the direction of impact parameter
than in its perpendicular direction on the transverse plane, which
are designated, respectively, as the $x$-direction and the $y$-direction.
The difference in the overlapping lengths along these two directions leads to difference in the pressure
gradient in these two directions, and as a result fluid velocities
along these two directions are different as well, with the fluid
velocity in the $x$-direction higher than the one in the $y$-direction.
To include such an anisotropic expansion of the fireball, we add a
term proportional to the eccentricity of the fireball to the
acceleration [Eq.~(\ref{acc})] as follows
\begin{eqnarray}
a_x&=&a_T (1+z\epsilon)\nonumber\\
a_y&=&a_T (1-z\epsilon),
\label{acc2}
\end{eqnarray}
where the eccentricity $\epsilon$ is defined as
\begin{eqnarray}
\epsilon=\frac{R_y-R_x}{R_y+R_x},\nonumber
\end{eqnarray}
and $z$ is a fitting parameter to take into account the different accelerations in and
perpendicular to the reaction plane in non-central collisions.
In the above, $R_x$ and $R_y$ are, respectively, the half length of
the fireball in the $x$-direction and in the $y$-direction on the
transverse plane. If the shape of the overlapping region is a perfect
ellipse, they then correspond to the semiminor and semimajor of the
ellipse, respectively. For simplicity, we transform the initial
shape of the fireball into a ellipse by keeping the same transverse
area and the same $R_x$ to $R_y$ ratio. The velocities in
the $x$-direction and in the $y$-direction are then, respectively,
\begin{eqnarray}
v_x^{i+1}=\frac{v_x^i+\Delta v_x}{1+v_x^i \Delta v_x}\nonumber\\
v_y^{i+1}=\frac{v_y^i+\Delta v_y}{1+v_y^i \Delta v_y},
\label{acc3}
\end{eqnarray}
where $\Delta v_x=a_x^i \Delta t$ and $\Delta v_y=a_y^i \Delta t$.
Assuming that the fireball keeps the elliptic shape on transverse
plane during expansion, the fluid velocity, which is assumed to
be along the radial direction, as a function of the azimuthal angle
$\phi$ is then
\begin{eqnarray}
v(\phi)=R^3(\phi) \bigg[\frac{v_x
\cos^2\phi}{R_x^3}+\frac{v_x\sin^2\phi}{R_y^3}\bigg],
\end{eqnarray}
where
\begin{equation}
R(\phi)=\bigg[\frac{\cos^2\phi}{R_x^2}+\frac{\sin^2\phi}{R_y^2}\bigg]^{-1/2}\nonumber
\end{equation}
is the radial distance from the center of the fireball to its
surface along the azimuthal angle $\phi$. Details in the derivation of
the above result are given in the Appendix. For the transverse
acceleration $a_T$ in Eq. (\ref{acc2}), it is obtained in the same
way as in central collisions. The constant $z$ in Eq. (\ref{acc2}) is
determined by requiring that the final $v_2$ of light hadrons,
given by the expectation value of $\cos (2\varphi)$ as
\begin{eqnarray}
&&v_2(p_T)=\frac{\int d\varphi \cos (2\varphi) (dN/dy d^2p_T)}{\int d\varphi (dN/dy d^2p_T)}\nonumber\\
&&=\frac{\int dA_T \cos (2\phi) I_2(p_T \sinh \rho/T) K_1(m_T \cosh \rho/T)} {\int dA_T I_0(p_T \sinh \rho/T) K_1(m_T \cosh \rho/T)},\nonumber\\
\label{v2}
\end{eqnarray}
reproduce the measured data.

\begin{figure}[h]
\centerline{
\includegraphics[width=8cm]{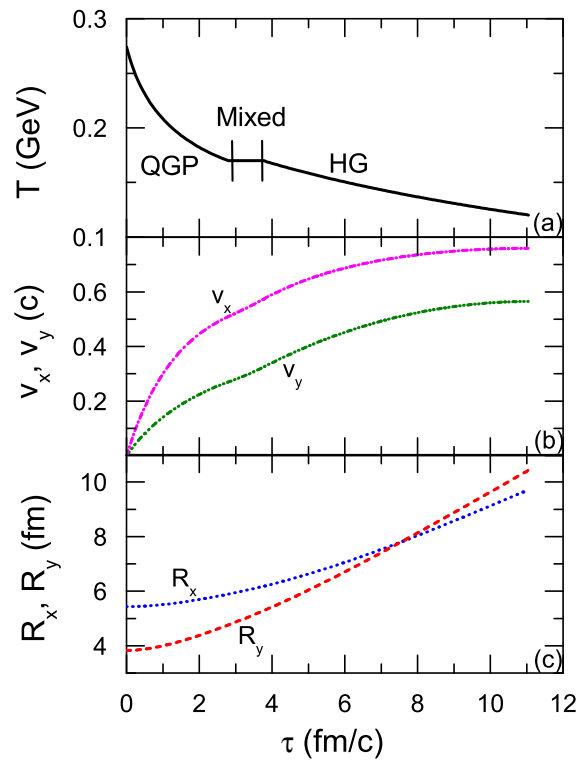}}
\caption{(color online) Time evolution of temperature (a), $v_x$
and $v_y$ (b), $R_x$ and $R_y$ (c) for Au+Au collisions at $\sqrt{s}=200$ GeV and impact parameter $b=$9 fm.}
\label{evolution}
\end{figure}

In Fig.~\ref{evolution}, we show the time evolution of the
temperature of the fireball and of $v_x$, $v_y$, $R_x$ and $R_y$
for Au+Au collisions at $\sqrt{s}=200$ GeV and impact parameter $b=$ 9 fm. These results are obtained with
the anisotropic constant $z=2.2$ in Eq. (\ref{acc2}) and
the inertia mass $M=108$ GeV in Eq. (\ref{acc}).
It is seen that the QGP phase ends at $t=2.8~{\rm fm}/c$, the
mixed phase lasts until $t=3.8~{\rm fm}/c$, and the
freeze-out temperature is reached at $t=11.1~{\rm fm}/c$. The flow velocities $v_x$ and $v_y$ are 0.51 $c$ and 0.27 $c$ at the phase transition to HG and reach 0.76 $c$ and 0.57 $c$ at
freeze-out, respectively. We note that the length of $R_x$ becomes slightly longer than that of $R_y$ at freeze-out.

\begin{figure}[h]
\centerline{
\includegraphics[width=8cm]{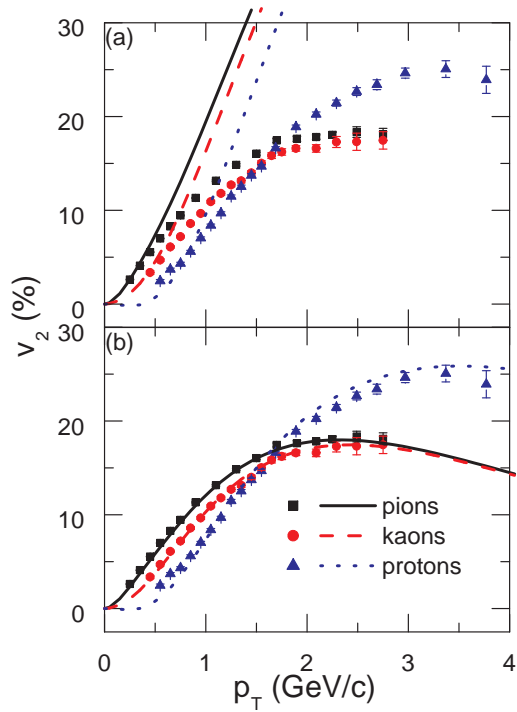}}
\caption{(color online) Elliptic flow $v_2$ of pions (solid
line), kaons (dashed line), and protons (dotted line) for Au+Au collisions
at $\sqrt{s}=200$ GeV and impact parameter $b=9$ fm before
the high $p_T$ correction (a) and after the correction (b).
Experimental data from collisions at 20-60\% centrality~\cite{Afanasiev:2007tv} are shown by rectangles, circles, and triangles for pions, kaons, and protons, respectively.}
\label{v2-light}
\end{figure}

The elliptic flow $v_2$ of pions (solid line), kaons (dashed line)
and protons (dotted line) at $b=9$ fm, obtained from Eq. (\ref{v2})
and using the results of Fig.~\ref{evolution}, are shown in
Fig.~\ref{v2-light}(a). Except in the low $p_T$ region, they are seen to deviate
from the experimental data from collisions at 20-60\% centrality
\cite{Afanasiev:2007tv}, shown by rectangles, circles, and triangles
for pions, kaons, and protons, respectively. The reason for
this deviation is because high-$p_T$ particles are not likely to be
completely thermalized as assumed in our fireball model.
To take into the non-equilibrium effect, we multiply $\exp[-C(p_T/n)]$ to $\Delta v=(v_x-v_y)/2$, where $C=1.14~{\rm GeV^{-1}}$ is a fitting parameter and $n$ is the number of constituent quarks in a hadron, as described in the Appendix.
Figure \ref{v2-light}(b) shows the resulting $v_2$ of light hadrons
after this high $p_T$ correction, and they are seen to
reproduce very well the experimental data at all $p_T$.  Since the
high $p_T$ correction factor modifies the Boltzmann distribution in Eq.~(\ref{number}),
it can be considered as an effective way to include the viscous effect on the fireball
expansion. In the quark-counting picture for hadron scattering, the ratio of the baryon to
the meson scattering cross sections is 3/2. This then leads to a smaller viscous effect on baryons than
mesons, which is consistent with the quark number dependence in the above high $p_T$ correction
factor.

\begin{figure}[h]
\centerline{
\includegraphics[width=8cm]{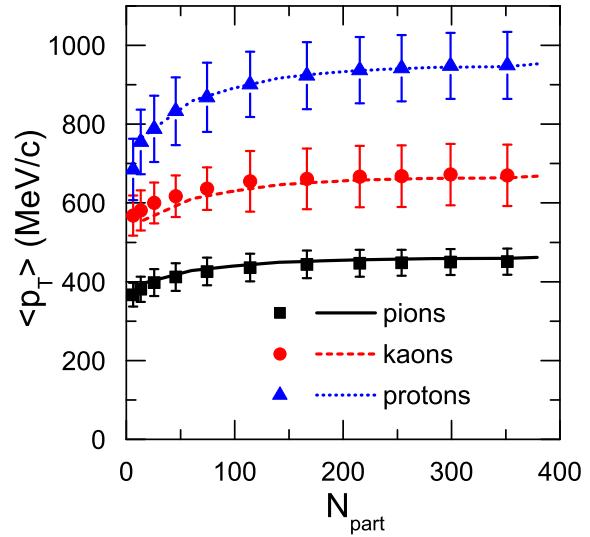}}
\caption{(color online) Participant number dependence of average transverse momenta of $\pi^+$ (solid line), $K^+$ (dashed line), and proton (dotted line) from the fireball model with corresponding experimental data shown by the rectangle, circle and triangle.} \label{resistance}
\end{figure}

The average transverse momenta of pions, kaons, and protons in the
above case of $b=9$ fm are $433$, $619$ and $874$ MeV/$c$,
respectively, which are comparable to measured values of 436$\pm$35,
655$\pm$54, and 901$\pm$88 MeV/$c$ in collisions at 30-40\% centrality as well as 426$\pm$35, 636$\pm$54 and 868$\pm$88 MeV/$c$ in collisions
at 40-50\% centrality for $\pi^+$, $K^+$ and proton, respectively
\cite{Adler:2003cb}. For other values of the impact parameter, it is
found that the inertia mass parameterized as $M=0.36
N_{\rm part}^{1.24}$ with $N_{\rm part}$ being the number of participants in a collision,
works well as shown in Fig. \ref{resistance}, where
the participant number dependence of the average
transverse momenta of $\pi^+$, $K^+$ and protons
obtained with the parameterized inertia mass are
compared with the experimental data.

\section{Nuclear modification factor of $J/\psi$}\label{raa}

The nuclear modification factor $R_{AA}$ of $J/\psi$ is a measure of its production
in heavy ion collisions relative to that in $p+p$ collisions multiplied
by the number of binary collisions. Its value is less than one
if there is a suppression and greater than one if there is an enhancement. Besides effects due to
the produced hot dense matter, the $R_{AA}$ of $J/\psi$ is also
affected by two other important effects, namely the Cronin effect
and the leakage effect.

\subsection{the Cronin effect}

Charmonium is produced mainly through the gluon-gluon fusion in
high-energy nuclear collisions. If a gluon in one nucleus is
scattered by the nucleon in the other colliding nucleus before it
produces the charmonium, the gluon obtains an additional transverse
momentum, and the charmonium produced from the gluons then also
acquires this additional transverse momentum. As a result, the
distribution of the transverse momentum of $J/\psi$ becomes
different from that in p+p collisions, and this
is called the Cronin effect. Taking into account of this effect,
the average transverse momentum of $J/\psi$ in A+A collisions is
modified as following:
\begin{eqnarray}
\langle p_T^2\rangle_{AA}=\langle p_T^2\rangle_{pp}+\frac{\langle\delta p_T^2\rangle}{\lambda_{gN}}\langle L_{gN}\rangle,
\end{eqnarray}
where $\langle p_T^2\rangle_{AA}$ and $\langle p_T^2\rangle_{pp}$
are average squared transverse momenta of charmonium in an A+A
collision and in a p+p collision, respectively; $\langle \delta
p_T^2\rangle$ is the average of added squared transverse momentum by
one gluon-nucleon collision, $\lambda_{gN}$ is the mean free path of
a gluon in uniform nuclear matter, $\langle L_{gN}\rangle$
is the average path length of the gluon in uniform nuclear matter before it produces a charmonium. The prefactor $\langle\delta p_T^2\rangle/\lambda_{gN}$ is fitted to the experimental data of p+A or d+A collisions and
$\langle L_{gN}\rangle$ is calculated with the Glauber model
according to
\begin{eqnarray}
\langle L_{gN}\rangle=\frac{\int d^2b d^2s dz_A dz_B L(\vec{b},\vec{s},z_A,z_B)K(\vec{b},\vec{s},z_A,z_B)}{\int d^2b d^2s dz_A dz_B K(\vec{b},\vec{s},z_A,z_B)},\nonumber\\
\end{eqnarray}
where
\begin{eqnarray}
&&L(\vec{b},\vec{s},z_A,z_B)=\int^{z_A}_{-\infty}dz \frac{\rho_A(\vec{s},z)}{\rho_o}+\int^{\infty}_{z_B}dz \frac{\rho_B(\vec{b}-\vec{s},z)}{\rho_o},\nonumber\\
\nonumber\\
&&K(\vec{b},\vec{s},z_A,z_B)=\rho_A(\vec{s},z_A)\rho_B(\vec{b}-\vec{s},z_B)\nonumber\\
&&\times \exp\bigg(-\sigma_{abs}\bigg[\int^\infty_{z_A}dz\rho_A(\vec{s},z)+\int^{z_B}_{-\infty} dz\rho_B(\vec{b}-\vec{s},z)\bigg]\bigg).\nonumber
\label{k}
\end{eqnarray}
In the above, $\vec{b}$ and $\vec{s}$ are the impact parameter and
transverse vector in configuration space, respectively; $z_A$ and
$z_B$ are longitudinal positions where $J/\psi$ is produced from the
center of nucleus A and from the center of nucleus B, respectively;
$\rho_A$ and $\rho_B$ are the distribution functions of nucleons in nucleus A and in nucleus B, respectively, and they are normalized to their mass numbers;
$\rho_o$ is the saturation density of nuclear matter. The function
$L(\vec{b},\vec{s},z_A,z_B)$ is the sum of the path length in
nucleus A of a gluon in nucleus B and the path length in nucleus B
of a gluon in nucleus A, which correspond to the first term and the
second term on the right hand side of $L(\vec{b},\vec{s},z_A,z_B)$,
respectively. The probability that a charmonium is produced at the
position $z_A$ from the center of nucleus A and $z_B$ from the
center of nucleus B is given by $K(\vec{b},\vec{s},z_A,z_B)$. The
exponential factor in $K(\vec{b},\vec{s},z_A,z_B)$ is due to the
effect of nuclear absorption or nuclear destruction as produced
$J/\psi$ can be absorbed or destroyed by nucleons with the
absorption cross section $\sigma_{abs}$ before it gets out of the
colliding nuclei. Notice that the integration range for the Cronin
effect is from $-\infty(\infty)$ to $z_A(z_B)$ and for nuclear
absorption from $z_A(z_B)$ to $\infty(-\infty)$ in nucleus A(B), as the Cronin effect is present before the production of the primordial
charmonium and the nuclear absorption happens after the creation of
the $J/\psi$. In the present work,
$L_{gN}$ is calculated at each point on the transverse plane as following:
\begin{eqnarray}
L_{gN}(\vec{b},\vec{s})=\frac{\int dz_A dz_B L(\vec{b},\vec{s},z_A,z_B)K(\vec{b},\vec{s},z_A,z_B)}{\int dz_A dz_B K(\vec{b},\vec{s},z_A,z_B)}. \nonumber\\
\end{eqnarray}
The Cronin effect thus depends on the transverse position at which a
$J/\psi$ is produced.

The transverse momentum distribution of $J/\psi$ in A+A collisions is related to that in p+p
collisions by
\begin{eqnarray}
f_{A+A}(\vec{p_T},\vec{b},\vec{s})&=& \frac{1}{\pi\delta p_T^2(\vec{b},\vec{s})} \int  d^2  p_T'  \exp\bigg[\frac{p_T'^{2}}{\delta p_T^2(\vec{b},\vec{s})}\bigg]\nonumber\\
&&~~~~~~~~\times f_{p+p}(|\vec{p_T}-\vec{p_T'}|),
\end{eqnarray}
where
\begin{eqnarray}
\delta p_T^2(\vec{b},\vec{s})=\frac{\langle \delta p_T^2\rangle}{\lambda_{gN}}L_{gN}(\vec{b},\vec{s})\nonumber
\end{eqnarray}
and $f_{p+p}(\vec{p_T})$ is the distribution function of $J/\psi$
transverse momentum in p+p collisions. For center of mass energy at
$\sqrt{s_{NN}}=$ 200 GeV, $f_{p+p}(\vec{p_T})=\alpha(1+p_T^2/\beta)^{-6}$ with $\beta=4.1$ ${\rm GeV^2}$ \cite{Adare:2006kf}, and we use
$\langle\delta p_T^2\rangle/\lambda_{gN}= 0.1$~GeV$^2$/fm and
$\sigma_{abs}=1.5$ mb for Au+Au collisions \cite{Zhao:2007hh}.

\subsection{the leakage effect}

In our previous work, it was assumed that charmonia produced inside
the fireball could not get out and charmonia produced outside could
not get into the fireball. In realistic situation, it is possible for
charmonia of high $p_T$ to escape from the fireball, leading
to the so-called leakage effect. To take into account this effect,
we consider the thermal decay width of a charmonium
inside the fireball, i.e., its transverse position ($x$ and $y$) satisfies
$[x/R_x(t)]^2+[y/R_y(t)]^2<1$,
\begin{eqnarray}
\Gamma(x,y,t)&=&\sum_i g_i \int\frac{d^3k}{(2\pi)^3}v_{\rm rel}(k)n_i(k,T) \sigma_i^{\rm diss}(k,T),\nonumber\\
\end{eqnarray}
where $i$ denotes the particle species that dissociates the
charmonium with $g_i$ and $n_i$ being its degeneracy factor and number
density, respectively; $v_{\rm rel}$ is the relative velocity between the
charmonium and the particle; $\sigma_i^{\rm diss}$ is the
dissociation cross section of charmonium by the particle. The
charmonium thermal decay width vanishes once it moves out of the fireball
when its transverse position satisfies $[x/R_x(t)]^2+[y/R_y(t)]^2>1$.
Furthermore, we set the charmonium decay width to infinite if it is produced
in a region with temperatures higher than its dissociation temperature.

\begin{figure}[h]
\centerline{
\includegraphics[width=8cm]{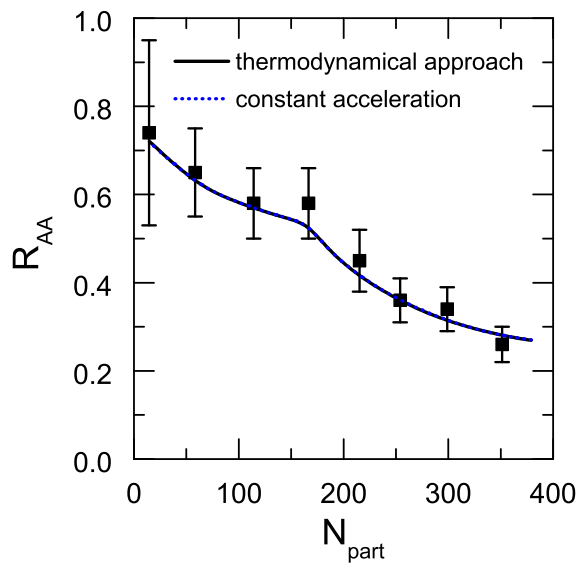}}
\caption{(color online) $R_{AA}$ of $J/\psi$ as a function of
the number of participants including both the Cronin and the leakage
effect. The dotted line is the $R_{AA}$ for constant acceleration
and the solid line is that for acceleration determined
thermodynamically, which are, however, almost indistinguishable.}
\label{Raa-np1}
\end{figure}

Results obtained with the two-component model described in Sec.~\ref{two}
for the $R_{AA}$ of $J/\psi$ as a function of the number of
participants including both the Cronin and the leakage effect are
shown in Fig.~\ref{Raa-np1}. The dotted line is the $R_{AA}$ with
constant acceleration of the fireball and the solid line is that
with the fireball acceleration determined thermodynamically. In the
latter case, the Cronin effect and the leakage effect are included.
As shown in Fig. \ref{comparison}(b), the lifetime of QGP is slightly longer in the former case than in the latter case, and this leads to a somewhat smaller $R_{AA}$ in the former case. Since the
Cronin effect only changes the transverse momentum of $J/\psi$, not
the number of $J/\psi$, it does not affect the $R_{AA}$ of $J/\psi$.
The leakage effect, on the other hand, may increase $R_{AA}$ or
decrease $R_{AA}$, depending on whether $J/\psi$ initially produced
outside the fireball can move into the fireball and on whether
$J/\psi$ initially produced inside the fireball can escape to the
outside. The net leakage effect on $R_{AA}$ is, however, found to be
small, and the difference between the results for constant acceleration and
those for acceleration determined thermodynamically is thus hardly seen in
Fig.~\ref{Raa-np1}.

\begin{figure}
\centerline{
\includegraphics[width=8cm]{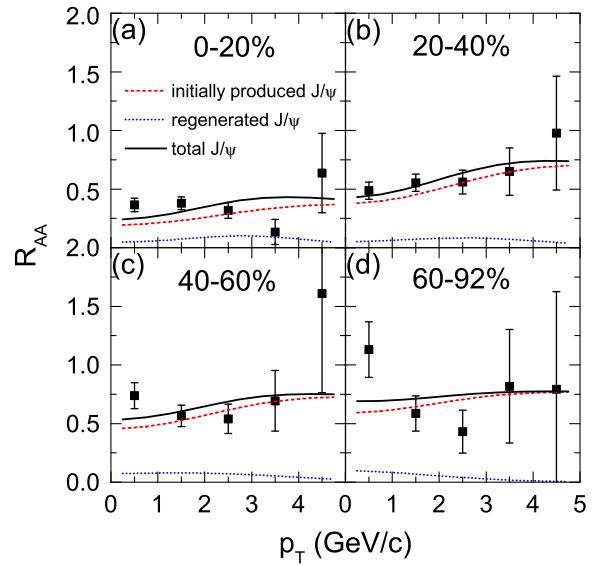}}
\caption{(color online) Nuclear modification factor $R_{AA}$
of $J/\psi$ as a function of transverse momentum in collisions of
0-20\% (a), 20-40\% (b), 40-60\% (c) and 60-92\% (d) centralities.
Dashed and dotted lines are, respectively, the $R_{AA}$ from initially produced $J/\psi$ and regenerated
$J/\psi$, and the solid line is the sum of the two.} \label{Raa-pt}
\end{figure}

Figure~\ref{Raa-pt} shows the $R_{AA}$ of $J/\psi$ as
a function of transverse momentum obtained with the fireball
acceleration calculated thermodynamically. It is seen that the
$R_{AA}$ of initially produced $J/\psi$, given by the dashed line,
increases with transverse momentum because of the Cronin effect. On
the contrary, the $R_{AA}$ of regenerated $J/\psi$ decreases with transverse
momentum at high $p_T$ as shown by the dotted line. We note that the
$R_{AA}$ of $J/\psi$ is slightly enhanced and suppressed at high and
low $p_T$, respectively.

\begin{figure}[h]
\centerline{
\includegraphics[width=8cm]{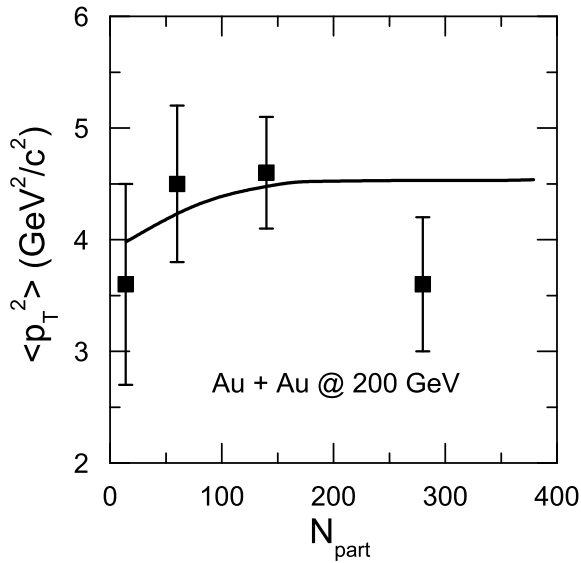}}
\caption{(color online) Average squared
transverse momentum of $J/\psi$ in Au+Au collisions at
$\sqrt{s_{NN}}=200$ GeV from the two-component model (solid line)
and experimental data (filled squares).} \label{pt2}
\end{figure}

In Figure~\ref{pt2}, we show the participant number dependence of the
average squared transverse momentum of $J/\psi$ in Au+Au collisions
from the two-component model. Our results reproduce the experimental
data for peripheral and mid-central collisions. For central
collisions, the two-component model overestimates the
measured average squared transverse momentum of $J/\psi$. One
possible reason for this is that we may have overestimated the
diffusion of charm quarks in the fireball. The number of charm
quarks produced in heavy-ion collision is proportional to the
number of binary collisions of nucleons, which is more concentrated
in the central region of the fireball. The number of light quarks
is, on the other hand, proportional to the number of participants.
Since the relative concentration of the number of binary collisions
to the number of participants is larger in central collisions
than in semi-central or peripheral collisions, charm quarks are
initially more concentrated in the central region than light quarks.
If the diffusion of charm quarks is slow, they would remain near the
central region even at hadronization and their transverse velocity
would then be smaller than those of light quarks.

\section{elliptic flow of $J/\psi$}\label{elliptic}

The elliptic flow $v_2$ of $J/\psi$ in the two-component model can
be different from that of light hadrons as a result of following two
effects. Firstly, $J/\psi$'s are produced from two different
mechanisms: the initial hard scattering and the regeneration, while
light hadrons are produced only through regeneration. Secondly, the
freeze-out temperature for $J/\psi$ could be different from that of
light hadrons. Since initially produced $J/\psi$'s have transverse
momenta that are azimuthally isotropic, they are expected to have
vanishing $v_2$ . However, the anisotropic expansion of the fireball
could give rise to nonzero $v_2$ for $J/\psi$. In noncentral
collisions, the size of the fireball is shorter in the $x$-direction
than in the $y$-direction. As a result, the survival rate of $J/\psi$
moving in the $x$-direction will be higher than that moving in
the $y$-direction. In this case, the $v_2$ of $J/\psi$ is positive,
because the number of $J/\psi$ moving in the $x$-direction is more than
that moving in the $y$-direction. On the contrary, the $v_2$ of
regenerated $J/\psi$ is negative at low $p_T$ for the following
reason. Assuming particles are locally thermalized, they then have a
thermal distribution in the local frame. Since the thermal
distribution is peaked at $p^2=2T(T+\sqrt{T^2+m^2})$, where $T$ and
$m$ are the temperature and mass of the particle, respectively,
more massive particles thus peak at higher momenta. If the thermalized
matter is boosted with a certain velocity, the peak is shifted to
higher momentum in the laboratory frame. In noncentral collisions,
the difference in the length of the overlapping region in the
$x$-direction and the $y$-direction leads to a difference in the
pressure gradient and as a result a difference in fluid velocity
along these two directions. Because the length of the overlapping
region is shorter in the $x$-direction, the fluid velocity is higher and
the peak of the momentum distribution is more shifted in that
direction. As the peak shifts to higher momentum, the number density
at momentum lower than the peak momentum becomes smaller than the
number density in the unshifted or less shifted momentum
distribution, while the number density at momentum higher than the
peak becomes larger. As a result, the number density of $J/\psi$
moving in the $x$-direction is smaller than that of $J/\psi$ moving in
the $y$-direction at small $p_T$, and this becomes opposite at high
$p_T$, leading thus to a negative $v_2$.

\begin{figure}[h]
\centerline{
\includegraphics[width=8cm]{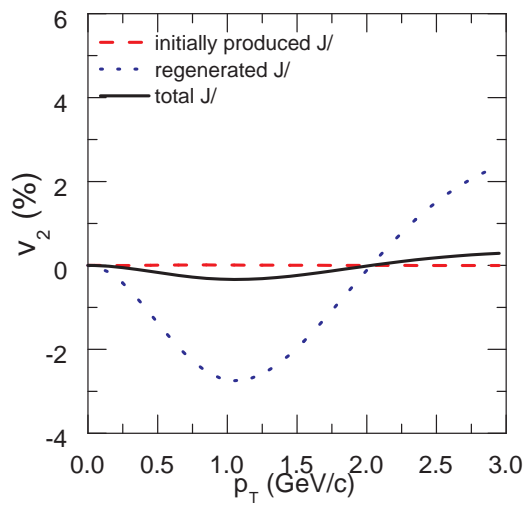}}
\caption{(color online) Elliptic flow $v_2$ of $J/\psi$ for collisions at
$b=9$ fm in the two-component model. The
dashed, dotted and solid lines are, respectively, the $v_2$ of
initially produced, regenerated and total $J/\psi$.}
\label{v2-jpsi1}
\end{figure}

Figure~\ref{v2-jpsi1} shows the results from the
two-component model for the $v_2$ of initially produced,
regenerated and total $J/\psi$ in Au+Au collisions at $\sqrt{s_{NN}}=200$ GeV
and $b=9$ fm. These results are obtained with the
high-$p_T$ correction to the flow velocity as in the case
for light hadrons. It is seen that the
$v_2$ of initially produced $J/\psi$ is essentially zero, whereas
the $v_2$ of regenerated $J/\psi$ has a minimum value of $-0.028$ at $p_T=1.1$
GeV. The $v_2$ of total $J/\psi$ is, however, slightly
negative, because only 12\% of $J/\psi$ are from regeneration for collisions at $b=9$ fm in the two-component model. We note that
the $v_2$ of $J/\psi$ has been recently measured by the PHENIX
collaboration for Au+Au collisions at 20-60\% centralities
\cite{Silvestre:2008tw,Krieg:2007bc} and it has a value of
-0.15$\pm$0.12 between $p_T=1$ and $2$ GeV,
which is much much more negative than our value.

\section{effect of higher-order pQCD corrections}\label{high}

In our two-component model, the screening mass between charm quarks
and the thermal decay widths of charmonia as well as the
relaxation factor of charm quarks in QGP are calculated in pQCD.
Specifically, the dissociation cross section of charmonium used for
determining its thermal decay width is calculated up to the
next-to-leading order, and the elastic cross section of charm quarks
for calculating the relaxation factor, and the screening mass are
calculated in the leading order. Since it is known
that pQCD is not reliable up to around $2~T_c$, corrections due to higher
orders are expected to be
important for the temperature regime considered in the present
study. The higher-order correction for the dissociation cross
sections of charmonia is expected to suppress the survival rate of
charmonia in hot dense nuclear matter while
the correction for the relaxation factor of charm quarks would
enhance the regeneration of charmonia. Therefore, there could be a
balance between the dissociation and the regeneration of charmonia in calculations including higher-order corrections. To include
higher-order corrections, we simply introduce in the present study
some multiplication factors to the dissociation cross sections
of charmonia and to the elastic cross section of charm quarks as following:
\begin{eqnarray}
\sigma_{J/\psi+q(g)\rightarrow c+\bar{c}+X}'&=& A \sigma_{J/\psi+q(g)\rightarrow c+\bar{c}+X}\nonumber\\
\sigma_{c+q(g) \rightarrow c+q(g)}'&=& B\sigma_{c+q(g) \rightarrow c+q(g)},
\end{eqnarray}
where $A$ and $B$ are arbitrary constants.

\begin{figure}[h]
\centerline{
\includegraphics[width=8cm]{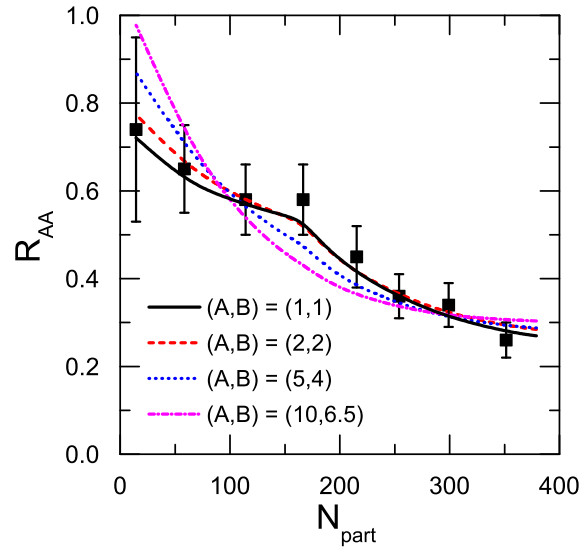}}
\caption{(color online) Nuclear modification factor $R_{AA}$
of $J/\psi$ as a function of the number of participants with
different values of the multiplicative parameteres (A,B)=(1,1),
(2,2), (5,4) and (10,6.5) for including the effect of
higher-order corrections in pQCD. Corresponding results are
shown by solid, dashed, dotted, and dot-dashed lines, respectively.}
\label{Raa2}
\end{figure}

In Fig.~\ref{Raa2}, the $R_{AA}$ of $J/\psi$ is shown as a function
of the number of participants. Solid, dashed, dotted, and dash-dotted
lines correspond, respectively, to results
obtained in the two-component model with (A,B)=(1,1), (2,2), (5,4),
and (10,6.5) for the multiplicative factors used to include
high-order effects in pQCD. These results show that higher-order corrections do not change
much the value of the $R_{AA}$ of $J/\psi$ even with a large correction
factor of ten. This is so because of the
balance between the dissociation and the regeneration of $J/\psi$ as
mentioned previously, with the suppression of $J/\psi$ caused by
enhanced dissociation compensated by the enhancement of regenerated
$J/\psi$. However, as the correction factor becomes large, the kink
in the $R_{AA}$ curve around $N_{\rm part}=170$ becomes less noticeable.
To understand this, we note that the kink in the curve is caused by the discontinuity
in the survival rate of $J/\psi$ in the region of the fireball where the
temperature is higher than the dissociation temperature and in the
region where the temperature is lower than the dissociation
temperature. As higher-order corrections become large, the
thermal decay width of $J/\psi$ increases and the survival rate of
$J/\psi$ in the low temperature region becomes comparable to the
survival rate in the high temperature region. As a result, the kink
disappears when higher-order corrections are large.

\begin{figure}[h]
\centerline{
\includegraphics[width=8cm]{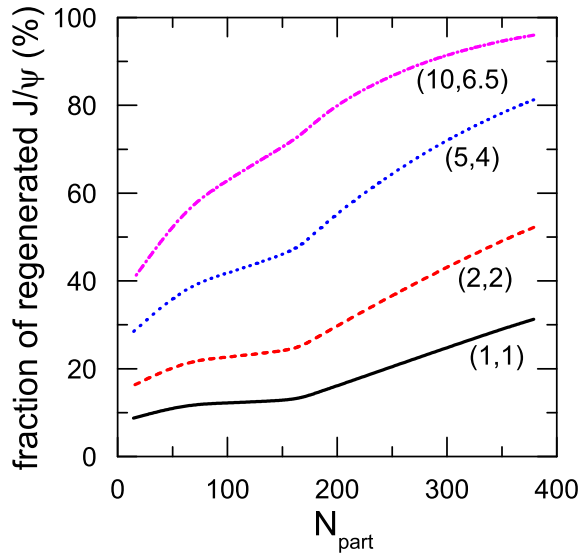}}
\caption{(color online) Fraction of regenerated $J/\psi$ among total $J/\psi$ in heavy-ion collisions. Solid, dashed, dotted,
and dot-dashed lines correspond, respectively, to multiplicative
factors (A,B)=(1,1), (2,2), (5,4) and (10,6.5) for higher-order
corrections.} \label{portion}
\end{figure}

Figure~\ref{portion} shows the fraction of regenerated
$J/\psi$ among total $J/\psi$ in heavy-ion collisions in the
two-component model. Without higher-order corrections (solid line),
the regenerated $J/\psi$ is about 31\% of total $J/\psi$ produced in
central collisions. Including higher-order corrections, the
percentage increases to 52, 81 and 96\%, respectively, for
multiplicative factors (A,B)=(2,2), (5,4) and (10,6.5), shown by
dashed, dotted, and dot-dashed lines, respectively. We note that the
faction of regenerated $J/\psi$ is smaller in semi-central and
peripheral collisions than in central collisions.

\begin{figure}
\centerline{
\includegraphics[width=8cm]{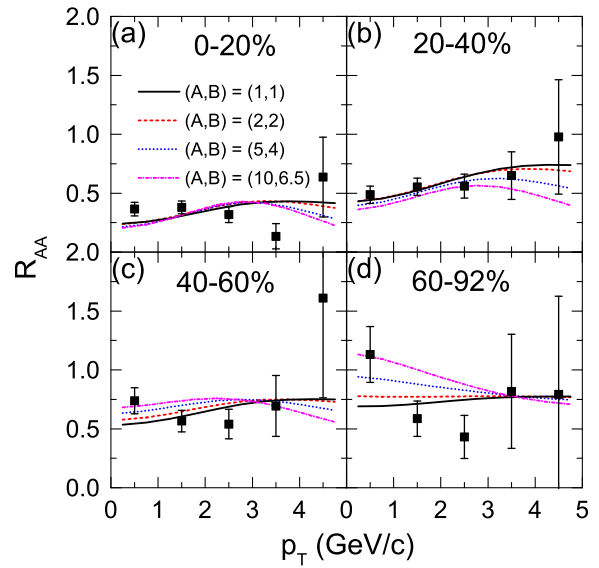}}
\caption{(color online) $R_{AA}$ of $J/\psi$ as a function of
transverse momentum in collisions of 0-20\% (a), 20-40\% (b),
40-60\% (c) and 60-92\% (d) centralities. Solid, dashed, dotted, and
dot-dashed lines correspond, respectively, to multiplicative factors
(A,B)=(1,1), (2,2), (5,4) and (10,6.5) for the higher-order
corrections.} \label{Raa-pt2}
\end{figure}

Figure~\ref{Raa-pt2} shows the $R_{AA}$ of $J/\psi$ as
a function of the transverse momentum in collisions of 0-20\% (a),
20-40\% (b), 40-60\% (c) and 60-92\% (d) centralities. The lines in each figure correspond,
respectively, to multiplicative factors (A,B)=(1,1),
(2,2), (5,4) and (10,6.5) for the higher-order corrections. As we
can see, $R_{AA}$ does not change much even with large higher-order
corrections as a result of the large transverse flow of regenerated
$J/\psi$. As shown in Fig. \ref{Raa-pt}, the $p_T$ dependence of the
$R_{AA}$ of initially produced $J/\psi$ is similar to that of
regenerated $J/\psi$ except at high $p_T$ where the $R_{AA}$ of
initially produced $J/\psi$ increases while that of regenerated
$J/\psi$ decreases. As a result, the $R_{AA}$ at high $p_T$ decreases
when higher-order corrections are included, and this seems to
contradict with the $p_T$ dependence of measured $R_{AA}$, which is
an increasing function at high $p_T$. Although higher-order
corrections are small at high $p_T$, they are relatively large at
low $p_T$, and this seems to reproduce the observed decreasing
$R_{AA}$ with increasing $p_T$, when $p_T$ is low, in peripheral
collisions as shown in Fig.~\ref{Raa-pt2}(d). This is mainly due to
the fact that in peripheral collisions, the time interval between
the thermalization of fireball and the freeze-out of $J/\psi$ is
short, so the transverse flow is not sufficiently developed to give
a large transverse flow velocity to the regenerated $J/\psi$.

\begin{figure}[h]
\centerline{
\includegraphics[width=8cm]{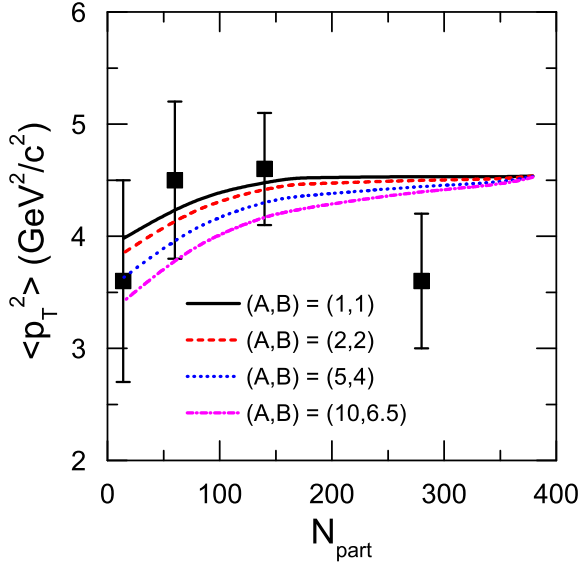}}
\caption{(color online) Average squared transverse momentum of
$J/\psi$ in Au+Au collision. Solid, dashed, dotted, and dot-dashed
lines correspond, respectively, to multiplicative factors
(A,B)=(1,1), (2,2), (5,4) and (10,6.5) for the higher-order
corrections.} \label{pt2b}
\end{figure}

In Fig.~\ref{pt2b}, we show the average squared transverse momentum
of $J/\psi$ before and after including higher-order corrections. It
is seen that higher-order corrections have very little effect on the
results.

\begin{figure}[h]
\centerline{
\includegraphics[width=8cm]{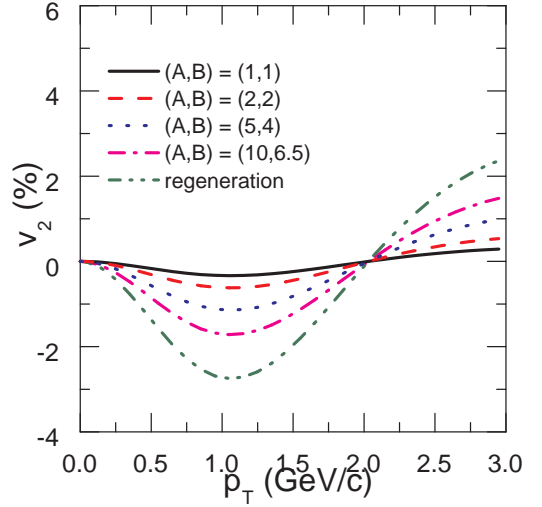}}
\caption{(color online) Elliptic flow $v_2$ of $J/\psi$ in collisions at
$b=9$ fm with higher-order corrections.
Solid, dashed, dotted, and dot-dashed lines correspond,
respectively, to multiplicative factors (A,B)=(1,1), (2,2), (5,4)
and (10,6.5) for the higher-order corrections. The double-dot-dashed
line is the $v_2$ of $J/\psi$ if they are totally from
regeneration.} \label{v2-jpsi2}
\end{figure}

Figure~\ref{v2-jpsi2} shows the $v_2$ of $J/\psi$ in collisions at
$b=9$ fm without and with higher-order corrections.
Different from the $R_{AA}$ of
$J/\psi$, the change in $v_2$ due to higher-order corrections is
significant. For increasing large correction factors listed in the
above, the fraction of regenerated $J/\psi$ among total $J/\psi$ is
12\%, 23\%, 42\% and 63\% respectively. This can be easily
understood from the fact that the $v_2$ of initially produced
$J/\psi$ is very small. In terms of the $v_2$ of initially produced
$J/\psi$ and that of regenerated $J/\psi$, the $v_2$ of total
$J/\psi$ is given by
\begin{eqnarray}
v_2&=&\frac{\int \cos(2\varphi)dN}{\int dN}=\frac{\int \cos(2\varphi)dN_R +\int \cos(2\varphi)dN_I}{\int dN_R
+\int dN_I}\nonumber\\
&\approx&\frac{\int \cos(2\varphi)dN_R}{\int dN_R
+\int dN_I}=\frac{\int \cos(2\varphi)dN_R}{\int dN_R
}\frac{N_R}{N},
\label{combination}
\end{eqnarray}
where $N$, $N_I$ and $N_R$ are, respectively, the total number of
$J/\psi$, the number of initially produced $J/\psi$ and that of
regenerated $J/\psi$. Equation (\ref{combination})
shows that the $v_2$ of total $J/\psi$ in heavy-ion collisions
can be approximated by the $v_2$ of regenerated $J/\psi$ multiplied
by its fraction among total $J/\psi$. The double-dot-dashed line in
Fig.~\ref{v2-jpsi2} is the $v_2$ of $J/\psi$ if all $J/\psi$ are
produced from regeneration. It is worthwhile to point out that even
in this extreme case, the $v_2$ of $J/\psi$ still only reaches the
upper error bar of experimental data, which has a value of $-0.03$
between $p_T=1$ and $2$ GeV \cite{Silvestre:2008tw,Krieg:2007bc}. If
the experimental data on $v_2$ of $J/\psi$ is correct, it then means
that the fraction of regenerated $J/\psi$ is not small even in
semi-central collisions.

\section{summary and conclusion}\label{summary}

We have investigated in the two-component model the $R_{AA}$ of
$J/\psi$ as a function of transverse momentum and as a function of
the number of participants as well as the $v_2$ of $J/\psi$ in relativistic heavy ion collisions.
Firstly, a schematic model was constructed to describe the expansion
of the fireball formed in these collisions.
The transverse acceleration of the fireball was assumed
to be proportional to the outgoing force given by the product of the
pressure of the hot dense matter inside the fireball and the surface
area of the fireball, and inversely proportional to an inertia mass,
which was taken to be a free parameter to fit the experimental data
on the measured transverse momentum spectra of light
hadrons. The anisotropic expansion in non-central collisions was
modeled by simply adding or subtracting an additional term
proportional to the eccentricity of the fireball to the acceleration
in $x$-direction or $y$-direction. The coefficient of the additional
term was determined such that the $v_2$ of light hadrons was
reproduced after including the high-$p_T$ correction to the flow velocity
due to the non-equilibrium effect. Next, the $R_{AA}$ and $v_2$ of $J/\psi$
were calculated with this schematic model. Both the Cronin effect
due to the initial transverse momentum broadening of gluons in
nuclei and the leakage effect due to the escape of initially
produced $J/\psi$ from produced hot dense matter were included, and
it was found that the $R_{AA}$ was suppressed at low $p_T$ and enhanced
at high $p_T$ by both effects. However, the leakage effect was much
smaller than the Cronin effect. For the kinetic freeze-out
temperature of $J/\psi$, it was taken to be the same as the chemical
freeze-out temperature, because the elastic cross section of
$J/\psi$ is much smaller than that of light hadrons, which normally
has a lower kinetic freeze-out temperature than the chemical
freeze-out temperature. Finally, higher-order corrections in pQCD
were included simply by multiplying separate constant factors to the
dissociation cross sections of charmonia and to the elastic cross
section of charm quarks. The former is related to the thermal decay
widths of charmonia and the latter to the relaxation factor of charm
quarks. Higher-order corrections were, however, not applied to the
screening mass. We found that higher-order corrections enhanced both
the dissociation of charmonia and the relaxation of charm quarks.
The former decreases the number of initially produced $J/\psi$ while
the latter increases the number of regenerated $J/\psi$. As a
result, the total number of $J/\psi$ was found to change very little
after including higher-order corrections. However, the fraction of
initially produced $J/\psi$ and that of regenerated ones were
affected significantly by higher-order corrections with larger
higher-order corrections leading to a larger fraction of regenerated
$J/\psi$ among total $J/\psi$. This is the reason why the $R_{AA}$
of $J/\psi$ as a function  of the number of participants was insensitive to the
higher-order corrections, and both the two-component model and the statistical model can describe
successfully the dependence of $R_{AA}$ on the number of
participants. Applying same higher-order corrections to the
transverse momentum dependence of the $R_{AA}$ of $J/\psi$, we found
that it was also insensitive to the fraction of regenerated $J/\psi$
among the total $J/\psi$ except at high $p_T$, as the small transverse momentum of regenerated
$J/\psi$ is compensated by the large transverse flow velocity. On
the other hand, the $v_2$ of $J/\psi$ is sensitive to the fraction
of regenerated $J/\psi$ because the $v_2$ of initially produced
$J/\psi$ is almost zero. If the elastic cross section of $J/\psi$ in
QGP is not small, the initially produced $J/\psi$ can then acquire
nonnegligible $v_2$ through scattering with thermal partons.
However, initially produced $J/\psi$ is expected to have a smaller
binding energy and larger radius because of the high temperature.
The loosely bound $J/\psi$ is expected to
be dissociated rather than scattered elastically when colliding with
thermal partons. For regenerated $J/\psi$,
they were assume to be produced from thermalized charm quarks in QGP. It is not clear yet how close charm
quarks are from thermal equilibrium in heavy-ion collisions.
Recent experimental data on non-photonic single electrons, which are
produced through the decay of heavy mesons, suggest that charm quarks
are thermalized significantly \cite{Adare:2006nq}. If charm quarks
are thermalized, resulting thermally produced $J/\psi$
would share their thermal properties and thus acquire appreciable $v_2$.
A possible future analysis would be to include the effect of abrupt decrease in mass of the charmonia near  $T_c$
\cite{Morita08}, as decreased mass will enhance the yield of regenerated $J/\psi$ and hence change its total contribution to $v_2$.
Presently, the $v_2$ of $J/\psi$ has large error bars in the experimental data. If
refined and precise data on the $v_2$ of $J/\psi$ become available,
they are expected to play an important role in discriminating between the two
production mechanisms for $J/\psi$, i.e., initial production and
regeneration.

\bigskip
\section*{Acknowledgements}
This work was supported in part by the U.S. National Science
Foundation under Grant No. PHY-0758115, the Welch Foundation
under Grant No. A-1358, the Korean Ministry of Education through the BK21 Program, and
the Korea Research Foundation under Grant No. KRF-2006-C00011.


\hfil\break
\appendix
\centerline{\bf \large Appendix}
\bigskip

Let the semimajor and the semiminor of an elliptic fireball produced
in heavy-ion collisions be $R_y$ and $R_x$, respectively, and
the fluid velocities along these two axes be $v_y$ and $v_x$,
respectively, at a certain time during the evolution of the
fireball. A short time $\Delta t$ later, the semimajor becomes
$R_y'=R_y+v_y \Delta t$ and the semiminor $R_x'=R_x+v_x \Delta t$.
Suppose that a certain point on the ellipse,
$(x/R_x)^2+(y/R_y)^2=1$, moves to a slightly expanded ellipse,
$(x/R_x')^2+(y/R_y')^2=1$, in radial direction keeping the same
polar angle $\phi$. Then the velocity of the point $v(\phi)$
satisfies
\begin{widetext}
\begin{eqnarray}
&&\frac{\bigg[ \{ R(\phi)+v(\phi)\Delta t \} \cos\phi
\bigg]^2}{\bigg[R_x+v_x \Delta t\bigg]^2} +\frac{\bigg[ \{
R(\phi)+v(\phi)\Delta t\} \sin\phi\bigg]^2}{\bigg[R_y+v_y \Delta
t\bigg]^2}=1, \label{ellipse}
\end{eqnarray}
where
\begin{eqnarray}
R(\phi)=\bigg[\frac{\cos^2\phi}{R_x^2}+\frac{\sin^2\phi}{R_y^2}\bigg]^{-1/2}\nonumber
\end{eqnarray}
is the distance from the center of the ellipse to the point on
the ellipse along the polar angle $\phi$. Expanding Eq.
(\ref{ellipse}) with respect to the infinitesimal time interval
$\Delta t$ gives
\begin{eqnarray}
\frac{\{R(\phi)\cos\phi\}^2}{R_x^2}+\frac{\{R(\phi)\sin\phi\}^2}{R_y^2}
+ \Delta t\bigg[v(\phi)
\bigg(\frac{2R(\phi)\cos^2\phi}{R_x^2}+\frac{2R(\phi)\sin^2\phi}{R_y^2}\bigg)
-\frac{2R(\phi)^2 v_x\cos^2\phi}{R_x^3}-\frac{2R^2(\phi)
v_y\sin^2\phi}{R_y^3}\bigg]=1,\nonumber
\end{eqnarray}
\end{widetext}
leading thus to the following expanding velocity of the surface of
the fireball
\begin{eqnarray}
v(\phi)=R^3(\phi) \bigg[\frac{v_x \cos^2\phi}{R_x^3}+\frac{v_y\sin^2\phi}{R_y^3}\bigg].
\label{flow-v}
\end{eqnarray}

Defining $v=(v_x+v_y)/2$ and $\Delta v=(v_x-v_y)/2$, Eq.~(\ref{flow-v}) can be rewritten as
\begin{eqnarray}
v(\phi)=R^3(\phi) \bigg[\frac{(v+\Delta v) \cos^2\phi}{R_x^3}+\frac{(v-\Delta v)\sin^2\phi}{R_y^3}\bigg].
\end{eqnarray}
Because high-$p_T$ particles are unlikely to follow the collective motion, both $v$ and $\Delta v$ can be functions of $p_T$. For simplicity, we modify only $\Delta v$ by multiplying it with the factor $\exp[-C(p_T/n)]$, where $C$ is a fitting parameter \cite{Oh:2009gx}.


\end{document}